\documentclass[12pt]{iopart}
\usepackage{epsfig}
\begin{document}

\title[Low-energy excitations in CeIn$_{3}$]
{Study of low-energy magnetic excitations in single-crystalline
CeIn$_{3}$ by inelastic neutron scattering}

\author{W. Knafo\dag , S. Raymond\dag, B. F{\aa}k\ddag, G.
Lapertot\dag, P.C. Canfield\S\, and J. Flouquet\dag}

\address{\dag\ CEA-Grenoble, DRFMC/SPSMS/MDN, 38054 Grenoble
Cedex, France}

\address{\ddag\ ISIS Facility, Rutherford Appleton Laboratory,
Chilton, Didcot, Oxon OX11 0QX, England}

\address{\S\ Ames Laboratory, Iowa State University, Ames Iowa 50011, USA}

\begin{abstract}

Inelastic neutron scattering experiments were performed on single
crystals of the heavy-fermion compound CeIn$_{3}$ for temperatures
below and above the N\'eel temperature, $T_N$.  In the
antiferromagnetically ordered phase, well-defined spin-wave
excitations with a bandwidth of 2 meV are observed.  The spin waves
coexist with quasielastic (QE) Kondo-type spin-fluctuations and
broadened crystal-field (CF) excitations below $T_N$.  Above $T_N$,
only the QE and CF excitations persist, with a weak temperature
dependence.

\end{abstract}



\submitted
{Printed: \today}


\section{Introduction}

A quantum critical point (QCP) occurs in the heavy fermion (HF)
compound CeIn$_{3}$ under a hydrostatic pressure of $P_{C}$ = 25
kbar. This corresponds to a transition at $T$ = 0 K from a
magnetically ordered to a disordered state.  It was found that at
least up to 23 kbar the ground state is antiferromagnetically
ordered with a propagation vector $\textbf{k} = (1/2,1/2,1/2)$. At
ambient pressure the N\'{e}el temperature is $T_{N}$ = 10.1 K and
the staggered magnetization is $m_{0}$ = 0.5 $\mu_{B}$
\cite{morin,benoit}. CeIn$_{3}$ becomes superconducting below
$T_{C}$ = 200 mK in the vicinity of $P_{C}$ \cite{mathur,knebel}.
More generally, the appearance of unconventional superconductivity
around $P_{C}$ in HF systems is supposed to be related to the
enhancement of magnetic fluctuations when $T_{N}\simeq0$.

CeIn$_{3}$ crystallizes in the cubic AuCu$_{3}$ structure, and hence
the magnetic fluctuations are expected to be three-dimensional (3D).
This is confirmed by resistivity measurements, which show a
$\Delta\rho = T^{3/2}$ non Fermi liquid (NFL) behavior at the critical
pressure, as expected for 3D fluctuations in spin-fluctuation theories
\cite{millis,moriya}.  Recently, a new family of HF compounds was
discovered \cite{thompson}, Ce$_{m}$T$_{n}$In$_{3m+2n}$, which is
composed of alternating layers of CeIn$_{3}$ and TIn$_{2}$ stacked
along the c-direction with T = Ir, Rh, or Co.  Their main interest arises
from the layered structure that is thought to induce 2D fluctuations
that would enhance the superconductivity more than 3D fluctuations
near $T_{C}$ \cite{monthoux}.  Indeed, CeCoIn$_{5}$ has a
superconducting state at ambient pressure below $T_{C}$ = 2.3 K, the
highest $T_{C}$ observed up to now in HF systems \cite{petrovic}.
These recent results motivated us to reinvestigate the common brick of
these quasi 2D compounds: the 3D heavy fermion CeIn$_{3}$.

Previous inelastic neutron scattering (INS) studies on powder samples
of CeIn$_{3}$ \cite{lawrence,murani} showed the existence of a crystal
field (CF) excitation at $\Delta_{CF}\simeq$ 12 meV with a large
broadening ascribed to the Kondo effect.  In this work, we present new
INS measurements on single crystals, which due to new flux growing
methods are now sufficiently large for such studies.  These
measurements show the existence of well-defined spin-wave excitations
in the magnetically ordered phase, which coexist with quasielastic
(QE) Kondo-type spin-fluctuations and the previously observed
broadened crystal-field excitations.

\section{Experimental details}

The single crystals used in this study were grown by the self flux
method \cite{canfield}.  A first set of neutron scattering
measurements was carried out on the time-of-flight chopper
spectrometer MARI at ISIS, using incident energies of 22 and 60.3 meV.
The corresponding energy resolution was approximately 2 and 5 meV,
respectively.  Because of the strong neutron absorption cross-section
of In, the single crystals were cut to a thickness of 2 mm and used in
transmission geometry.  A total of 30 crystals with total weight of 13
g were aligned and glued to a thin aluminium plate, which was mounted
in a closed-cycle refrigerator.  The mosaicity of the assembly was
approximately 6$^\circ$.  Measurements were performed in the (hhl) plane,
with an angle between the incident wavevector and the [001] direction
of 2.5 and 30$^\circ$.

A second set of neutron scattering measurements was made on the
triple-axis spectrometer IN22 at the ILL (Grenoble, France).
Pyrolytic graphite was used for the vertically focusing monochromator
and the horizontally focusing analyzer.  A PG filter in the scattered
beam was used to suppress higher order contamination, and final
neutron energies of 8.05 and 14.7 meV were used, giving an energy
resolution of 0.5 and 1.0 meV, respectively.  Two thin plates of the
assembly used on MARI were aligned on an aluminum support and mounted
in a helium flow "orange" cryostat.  The total mass of the sample was
about 1 g and the mosaicity 0.5$^\circ$.  The (hhl) scattering plane was
investigated.

\section{Results}
\subsection{Crystal-field excitation and quasielastic line}

For the two crystals orientations used on MARI and at temperatures
of $T$ = 5, 15 and 150 K, we recorded $(\textbf{Q},E)$ intensity
maps for 0$ < Q < $10 \AA$^{-1}$ and $E < 30$ meV, where
\textbf{Q} is the wavevector transfer and $E$ is the energy
transfer.  Despite strong phonon contamination from both the
CeIn$_{3}$ sample and the Al support, we observed an excitation
around 12 meV which was independent of the crystal orientation and
whose intensity decreased with $Q$. The temperature dependence of
the 12 meV excitation obtained for $Q < 2$ \AA$^{-1}$ is shown in
figure 1.  The excitation shows only a very slight softening and
broadening as the temperature goes from the ordered state at $T=5$
K to the paramagnetic state at $T=15$ K. At much higher
temperatures, $T$ = 150 K, the excitation is partly suppressed.
Figure 2 shows the wavevector dependence of the excitation
spectrum at $T=5$ K. The peak at 12 meV observed at low $Q$
(integration over $0<Q<2$ \AA$^{-1}$) disappears at higher wave
vectors (integration over $7.5<Q<7.8$ \AA$^{-1}$), reflecting its
magnetic nature.  The contribution around 20 meV, seen only at
high $Q$'s, is due to phonons.  The $Q$ and $T$ dependence of the
12 meV excitation, as illustrated in Figs.  1 and 2, clearly show
its magnetic origin, and corresponds to the CF excitation observed
in powder samples by Lawrence et al. \cite{lawrence} and Murani et
al. \cite{murani}.

The cubic environment of Ce$^{3+}$ in CeIn$_{3}$ can be treated in
terms of Stevens operators by the following CF Hamiltonian: $H_{CF} =
B_{4}(O_{4}^{0}+5O_{4}^{4})$ \cite{lea}.  This Hamiltonian splits the
$J = 5/2$ spin-orbit level of Ce$^{3+}$ into a $\Gamma_{8}$ quartet
and a $\Gamma_{7}$ doublet.  In the case of CeIn$_{3}$, susceptibility
measurements indicates that $B_{4} > 0$ \cite{buschow}, implying that
the ground state is $\Gamma_{7}$.  This has been confirmed by
polarized neutron scattering measurements, which also reveals an
admixture of $\Gamma_{8}$ to the ground state $\Gamma_{7}$ in the
ordered phase \cite{boucherle}.  The excitation observed here at 12
meV corresponds then to the $\Gamma_{7} \rightarrow \Gamma_{8}$
excitation.

In addition to the 12 meV CF excitation, it is clear from the data
that there is an additional quasielastic (QE) component to the
scattering.  The low-$Q$ data at $T=5$ K were therefore analyzed
using three components: an incoherent elastic peak, a quasielastic
spin-fluctuation contribution, and a crystal-field excitation.
Since the different contributions partly overlap, we fixed the
width (half width at half maximum, HWHM) $\Gamma_{QE}$ of the QE
contribution to  $k_{B}T_{K}$ = 0.86 meV where $T_{K}$ = 10 K is
the Kondo temperature. The results of the analysis at $T$ = 5 K
give a crystal-field splitting of $\Delta_{CF} = 10.6\pm0.2$ meV
and a broadening of the CF levels (HWHM) of $\Gamma_{CF} =
5.7\pm0.2$ meV.  No significant difference in the  values of the
CF excitations were found for the different orientations of the
crystal. At $T=15$ K, the resulting CF splitting is $\Delta_{CF} =
9.4\pm0.2$ meV and its width is $\Gamma_{CF} = 6.0\pm0.2$ meV. The
slight softening observed of $\Delta_{CF}$ is related to the
splitting of the CF levels below $T_{N}$.

\subsection{Spin wave}

The dispersive mode that appears below $T_{N}$ was first observed
using the MARI chopper spectrometer.  Because of the 3D character of
this mode, the study of its dispersion was continued using the IN22
triple-axis spectrometer.  Energy scans up to 6 meV energy transfer
were measured for different values of the wavevector transfer
$\textbf{Q}$.  Figure 3 shows spectra of CeIn$_{3}$ at $\textbf{Q} =
(-1.2,-1.2,0.5)$ for $T = 1.5$ and 15 K. This corresponds to the
reduced wavevector $\textbf{q} = \textbf{Q} - \textbf{$\tau$} -
\textbf{k} = (0.3,0.3,0)$, where $\textbf{$\tau$} = (-1,-1,1)$ is a
reciprocal lattice vector and $\textbf{k} = (-1/2,-1/2,-1/2)$ is a
magnetic propagation vector (wavevectors are expressed  in
reciprocal lattice unit with 1 r.l.u. = 2 $\pi/a$, where $a= 4.689$
\AA\ is the lattice parameter).  Figure 3 clearly shows that the
excitation present at $T < T_{N}$ disappears for $T > T_{N}$.  For
each wavevector $\textbf{q}$, the excitation was fitted by a Gaussian
whose width corresponds to the instrumental resolution.  The strong
dispersion of this excitation in the three main directions of the
crystal is shown in figure 4. Since the direction of the ordered
moment in CeIn$_3$ is unknown and our  sample is in a multidomain
state, it is not possible to determine the polarization of
the excitation; it will be assumed to be transverse as for a
conventional spin wave.

Supposing that the CF energy is of the same order as the exchange
interaction, it seems appropriate to treat the spin wave using the
so-called molecular-field random-phase approximation
\cite{buyers}.  Writing the exchange Hamiltonian as
\begin{equation}
\ H_{ex}=-\sum_{i,j}I_{ij}\textbf{J}_{i}\textbf{J}_{j}
\end{equation}
gives the general dispersion relation
\begin{equation}
E(\textbf{q})=\{[\Delta-M^{2}I(\textbf{q})]
[\Delta-M^{2}I(\textbf{q+k})]\}^{1/2}, \label{eq1}
\end{equation}
where $\textbf{k}$ is the propagation vector of the ordered state,
$I(\textbf{q})$ the Fourier transform of the exchange interaction
$I_{ij}$, and $\Delta$ the splitting of the $\Gamma_{7}$ doublet
into $\Gamma_{7,1}$ and $\Gamma_{7,2}$ due to the molecular and
anisotropy fields.  The transition matrix element between the
ground state $\Gamma_{7,1}$ and the first excited state
$\Gamma_{7,2}$ is given by
$M^{2}=|<\Gamma_{7,2}|J^{\alpha}|\Gamma_{7,1}>|^2$ ($\alpha=$ + or
-).  Considering only the nearest neighbor exchange $I_{1}$ [along
(1,0,0)] and the next nearest neighbor exchange $I_{2}$ [along
(1,1,0)], the exchange interaction is \begin{eqnarray}
I(\textbf{q})& =&2I_{1}[cos(2\pi
q_{x})+cos(2\pi q_{y})+cos(2\pi q_{z})]\nonumber \\
&+& 2I_{2}[cos(2\pi (q_{x}-q_{y}))+cos(2\pi (q_{y}-q_{z}))+cos(2\pi(q_{z}-q_{x})) \nonumber \\
&&\:\,+ cos(2\pi (q_{x}+q_{y}))+cos(2\pi
(q_{y}+q_{z}))+cos(2\pi(q_{z}+q_{x}))].
\end{eqnarray}
In a two sub-lattice picture constituted of alternating (1,1,1)
planes where the spins are up or down, $I_{1}$ would correspond to
the antiferromagnetic exchange between spins from different
sublattices and $I_{2}$ would correspond to the ferromagnetic
exchange between spins from the same sublattice.  A global fit to
the dispersion curves of equation (\ref{eq1}) is shown in figure
4.  It gives the exchange terms $M^{2}I_{1}= -0.354 \pm 0.040$ meV
and $M^{2}I_{2}= 0.028 \pm 0.013$ meV and the $\Gamma_{7}$
splitting $\Delta = 2.812 \pm 0.480$ meV. The corresponding
spin-wave gap is : $\delta = E(\textbf{0}) =$ 1.28 meV.

The exchange energy can be expressed by
$E_{ex}=\frac{2J(J+1)}{3}I(\textbf{0})$. Taking the wave functions
of $\Gamma_{7,1}$ and $\Gamma_{7,2}$ when there is neither
exchange nor distortion \cite{wang} gives us an approximated
square transition matrix element $M^{2}$ = 2.78 thus
$I(\textbf{0})=-0.64\pm0.18$ meV. With $J=5/2$, we obtain
$E_{ex}\simeq3.71$ meV. The ratio $\Delta_{CF} / |E_{ex}| \simeq
3$ permits then to conclude that the CF energy is of the same
order than the exchange interaction that justifies the treatment
of the spin wave using the molecular-field random-phase
approximation.

\section{Discussion}
\subsection{Kondo temperature}

We have analyzed the Q-independent excitations using two
contributions, one corresponding to quasielastic scattering and
the other to a crystal-field transition.  Since these excitations
overlap with each other and with the incoherent peak (see figure
2), the energy, width, and weight of the different contributions
extracted from the data depend on the model used for the fitting.
This is illustrated by the scatter in the estimates of
$\Delta_{CF}$, $\Gamma_{CF}$, and $\Gamma_{QE}$ found in the
literature \cite{lawrence,murani,gross,lassailly}.  At low
temperatures, the QE excitation corresponds to the spin
fluctuations of the $\Gamma_{7}$ electronic level, which are
attributed to the Kondo effect.  Its width (HWHM) $\Gamma_{QE}$
corresponds to the Kondo temperature.  A Kondo temperature of
$T_{K}$ = 10 K at a pressure of $P$ = 19 kbar was obtained in
$^{115}$In-NQR measurements by Kawasaki et al.  \cite{kawasaki}.
It is likely that below this pressure the anomaly at $T_{N}$ in
the $1/T_{1}$ temperature variation masks the $T_{K}$ anomaly.  We
assume therefore that $T_{K}$ has a very slight variation for
pressures below $P$ = 19 kbar, so that it is only slightly smaller
than 10 K at ambient pressure.  Resistivity measurements give a
Kondo scattering maximum at $T_{M}\simeq$ 50 K \cite{knebel},
which corresponds to the high Kondo temperature $T_{K}^{h} =
\sqrt[3]{T_{K}(\Delta_{CF}/k_{B})^{2}}$ \cite{hanzawa}. INS gives
$\Delta_{CF}=$ 10.6 meV which with $T_{K}^{h}=$ 50 K leads to
$T_{K}\simeq$ 8 K. It is then reasonable to approximate the Kondo
temperature by $T_{K}\simeq$ 10 K in CeIn$_{3}$ at ambient
pressure.  For the INS measurement carried out at $T=$ 5 K
$<T_{K}$, we choose consequently to fix the QE width to
$\Gamma_{QE}=k_{B}T_{K}=$ 0.86meV. The fact that $\Gamma_{CF}=$
5.7 meV is relatively large compared to $\Gamma_{QE}=$ 0.86 meV is
due to the larger spin fluctuations associated with the
$\Gamma_{8}$ first excited level.  In fact, the CF transition
$\Gamma_{7} \rightarrow \Gamma_{8}$ has a width corresponding to
the convolution of the $\Gamma_{7}$ and $\Gamma_{8}$ widths
$\Gamma_{QE}^{\Gamma_{7}}=\Gamma_{QE}$ and
$\Gamma_{QE}^{\Gamma_{8}}$. We can approximate
$\Gamma_{CF}\simeq\Gamma_{QE}^{\Gamma_{7}}+\Gamma_{QE}^{\Gamma_{8}}$
and thus estimate that
$\Gamma_{QE}^{\Gamma_{8}}\simeq\Gamma_{CF}-\Gamma_{QE}=$ 4.84 meV,
which is effectively 5 times larger than $\Gamma_{QE}$.

\subsection{Anisotropy}

Even if the CF level is broad, it seems that the propagating spin
waves can be described by considering $\Gamma_{7}$ as an
"isolated" doublet split by the anisotropy and exchange fields.
Since this doublet originates from a sixfold degenerate $J=5/2$
level, an anisotropy gap is expected in the spin-wave spectrum. We
recall that in the isotropic case, the spin-wave dispersion has no
gap, since only the exchange is responsible for the splitting
$\Delta$.

The gap $\delta$ of the spin wave excitation corresponds to its
minimum energy that is obtained at $\textbf{q}=(0,0,0)$. Since
this excitation overlaps with the incoherent elastic scattering
that reflects the experimental resolution and with the
quasielastic contribution of the spin fluctuations, the gap was
not correctly obtained in the present neutron experiment. The
given value of $\delta$ corresponds to the best fit to the data
using the spin-wave model described above. An attempt to directly
observe the gap using a smaller incident energy (and hence better
resolution) on the cold triple-axis spectrometer IN12 (ILL) failed
because of the strong indium absorption cross-section. A
determination of the energy gap was also made by low-temperature
specific-heat measurements which exhibited a spin wave
contribution that leads to a gap $\delta = 0.69$ meV \cite{us}.
This suggests that the gap energy of 1.28 meV deduced from our INS
measurements is somewhat overestimated.

The result of our spin-wave analysis is in good agreement with
calculations made by Wang et al. for Ce$^{3+}$ ions under cubic
crystal field and effective exchange field \cite{wang}. They
consider a molecular field along the c-axis, which can be
approximated by $g\mu_{B}H_{m} = k_{B}T_{N}$, and calculate the
level splitting as a function of $g\mu_{B}H_{m}/B_{4}$. Knowing
that for CeIn$_{3}$ $T_{N}$ = 10.1 K, $\Delta_{CF} = 10.6$ meV and
hence $B_{4} = 2.94\,10^{-2}$ meV ($\Delta_{CF} = 360\, B_{4}$),
we have $g\mu_{B}H_{m} = 8.71\,10^{-1} $meV $\simeq 30 \,B_{4}$.
For this value of $g\mu_{B}H_{m}/B_{4}$, Wang et al. calculate a
$\Gamma_{7}$ splitting of about $40 \,B_{4} \simeq 1.2$ meV.
Although we suppose that in our case the molecular field is not
along the c-axis (see below), we believe that Wang's calculations
give the right order of the splitting $\Delta$.

There are several indications for that the moments are not along
the c-axis.  If the CF alone imposes the easy magnetization axis,
the positive sign of $B_{4}$ in the CF Hamiltonian implies that
the moments are along the [1,1,1] direction in the ordered phase.
This is also supported by nuclear quadrupolar resonance
measurements \cite{kohori}. Although it is difficult to determine
the moment direction of a cubic compound by neutron diffraction,
some recent measurements also suggest that the moments are not
along the c-axis \cite{us}.

\subsection{Comparison with CePd$_{2}$Si$_{2}$}

CeIn$_{3}$ seems to have a behavior very similar to the HF
compound CePd$_{2}$Si$_{2}$ which is antiferromagnetically ordered
at ambient pressure and has the same kind of excitation spectrum
with coexisting QE, CF, and spin-wave excitations \cite{vandijk}.
The characteristic energies are very close and, moreover,
unconventional superconductivity appears at a similar external
pressure \cite{mathur}. A summary of the characteristic physical
quantities of those two compounds is given in Table 1. A focus on
the few differences between those two compounds can be made as
following. Firstly CePd$_{2}$Si$_{2}$ is tetragonal while
CeIn$_{3}$ is cubic. Some characteristics such as the CF splitting
or the susceptibility anisotropy can be directly linked to
anisotropy effects that are resulting from those lattice
structures. Then, the characteristic quantities $T_{C}$ and
$\gamma$  are twice bigger in CePd$_{2}$Si$_{2}$ than in
CeIn$_{3}$, that could traduce an enhancement of the spin
fluctuations in the tetragonal CePd$_{2}$Si$_{2}$. As a supplement
to the studies of quasi 2D Ce$_{m}$T$_{n}$In$_{3m+2n}$ compounds
where $T_{C}$ can reach more than 2 K, an alternative way to study
the influence of anisotropy in strongly correlated systems could
consequently remain in a detailed comparison between cubic
CeIn$_{3}$ and tetragonal CePd$_{2}$Si$_{2}$.

\begin{table}
\caption{Characteristic physical quantities of CeIn$_{3}$ and
CePd$_{2}$Si$_{2}$. Data are taken from \cite{mathur},
\cite{knebel}, \cite{us}, \cite{vandijk}, \cite{grier},
\cite{demuer}, \cite{steeman}, \cite{grosche}.}
\begin{center}
\item[]\begin{tabular}{|c|c|c|} \hline
 & CeIn$_{3}$ & CePd$_{2}$Si$_{2}$\\
\hline
$T_{N}$&10.1 K&10 K \\
\hline
$\textbf{k}$&(1/2,1/2,1/2)&(1/2,1/2,0)\\
\hline
$m_{0}$& 0.50 $\mu_{B}$& 0.62 $\mu_{B}$ \\
\hline
$\delta$&0.69 meV & 0.83 meV \\
\hline
Bandwidth&$\sim$ 2 meV &$\sim$ 2 meV \\
\hline
 $\theta_{P}$ &-50 K  &-47 K $\,^{*}$ \\
\hline
$T_{K}$&10 K &10 K  \\
\hline
$\Delta_{CF}$&123 K&220 and 280 K \\
\hline
$P_{C}$&26.5 kbar &28.6 kbar \\
\hline
$T_{C}$&200 mK &430 mK \\
\hline
$\gamma\,^{**}$ &130 mJ.mol$^{-1}$.K$^{-2}$  &250 mJ.mol$^{-1}$.K$^{-2}$\\
\hline
\end{tabular}
\end{center}
$\,^{*}$ : DC susceptibility measurement along a-axis

$\,^{**}$ : $\gamma=C/T$ at low T
\end{table}

\section{Conclusion}

We have performed the first inelastic neutron scattering
measurements on single crystals of CeIn$_{3}$.  The combined use
of time-of-flight and triple-axis spectrometers allowed us to
study both local and dispersive low-energy excitations.  We found
the coexistence below $T_{N}$ of a spin-wave with a quasielastic
component and a crystal-field excitation.  The CF levels are
substantially broadened by Kondo-type spin fluctuations, which
indicates the proximity to an intermediate valence state where the
CF collapse. Despite the smearing of the CF levels, the
observation of spin waves means that the ground state level in
CeIn$_{3}$ is also relatively well-defined. The challenge is then
to understand how these excitations evolve on going toward a QCP
where the hierarchy between the Kondo effect and the RKKY coupling
is reversed while the CF should be close to a collapse.

\section*{Acknowledgments}

We acknowledge the help of K. Mony for sample preparation and
X-ray characterization. We thank also A.P. Murani for many useful
and stimulating discussions.

\begin{figure}[h]
    \centering
    \epsfig{file=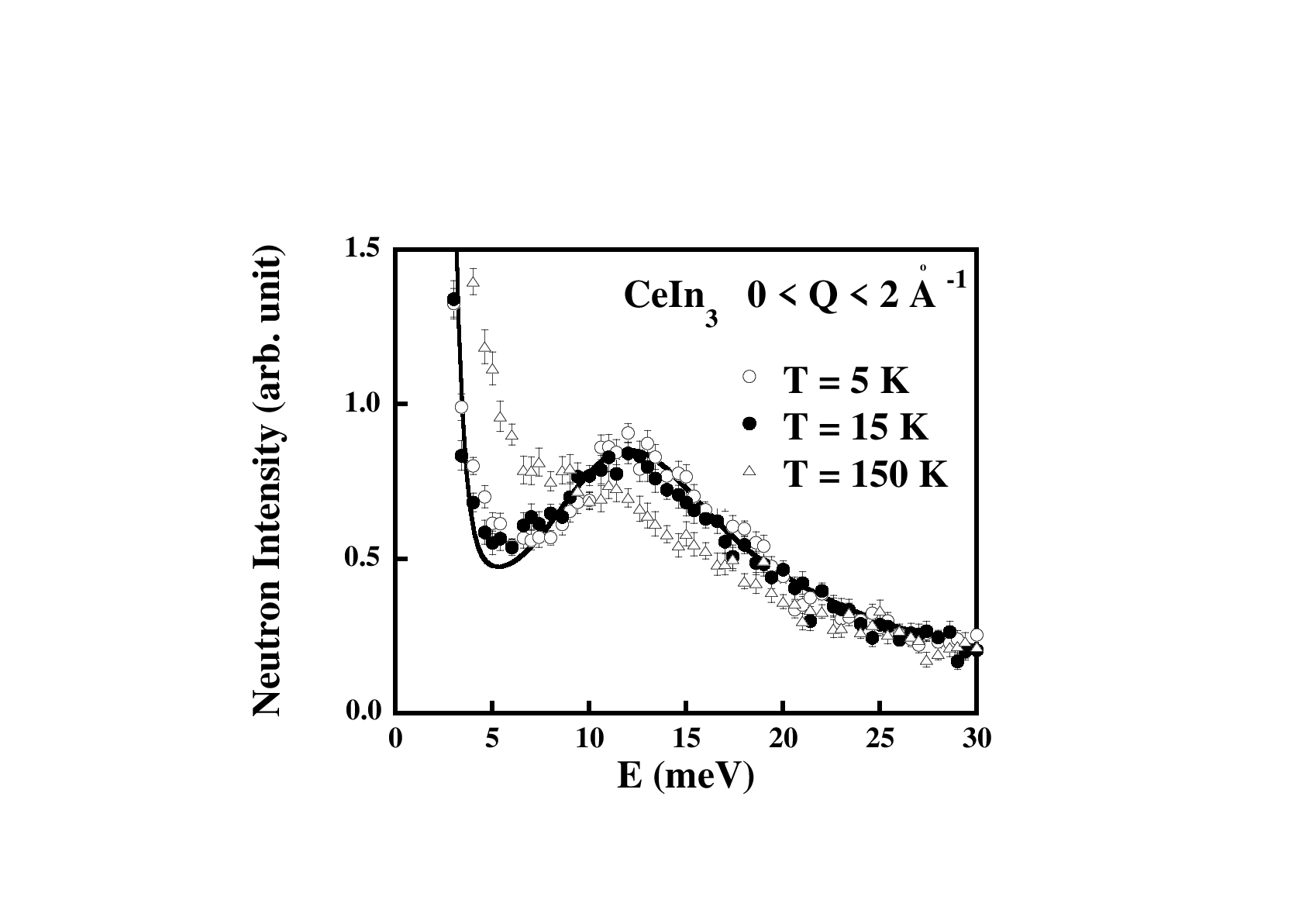,height=150mm,angle=0}
    \caption{Energy scans at different temperatures obtained on
    MARI spectrometer using an incident energy of 60.3 meV and
    integrating over $Q$ values between 0 and 2 \AA $^{-1}$. The
    solid line shows the total fit for $T$ = 5 K.}
    \label{fig1}
\end{figure}

\begin{figure}[h]
    \centering
    \epsfig{file=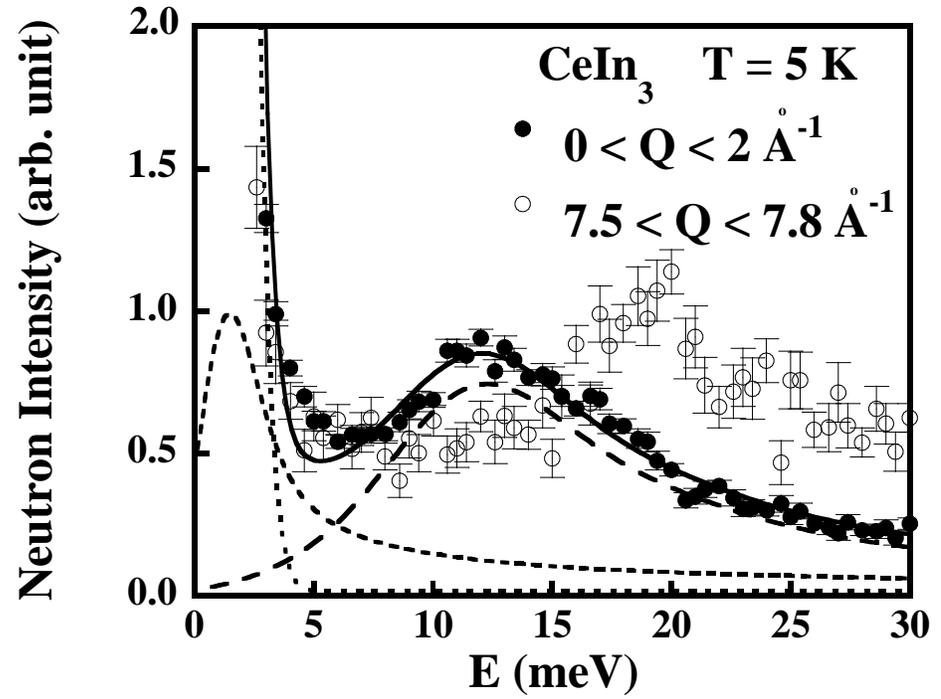,height=150mm,angle=0}
    \caption{Energy scans at different wavevectors obtained on the
    MARI spectrometer using an incident energy of 60.3 meV at $T=$ 5
    K. The long dashed, short dashed, and dotted lines show the CF,
    QE, and incoherent contributions of the small Q spectrum,
    respectively, while the solid line shows the total fit.}
    \label{fig2}
\end{figure}

\begin{figure}[h]
    \centering
    \epsfig{file=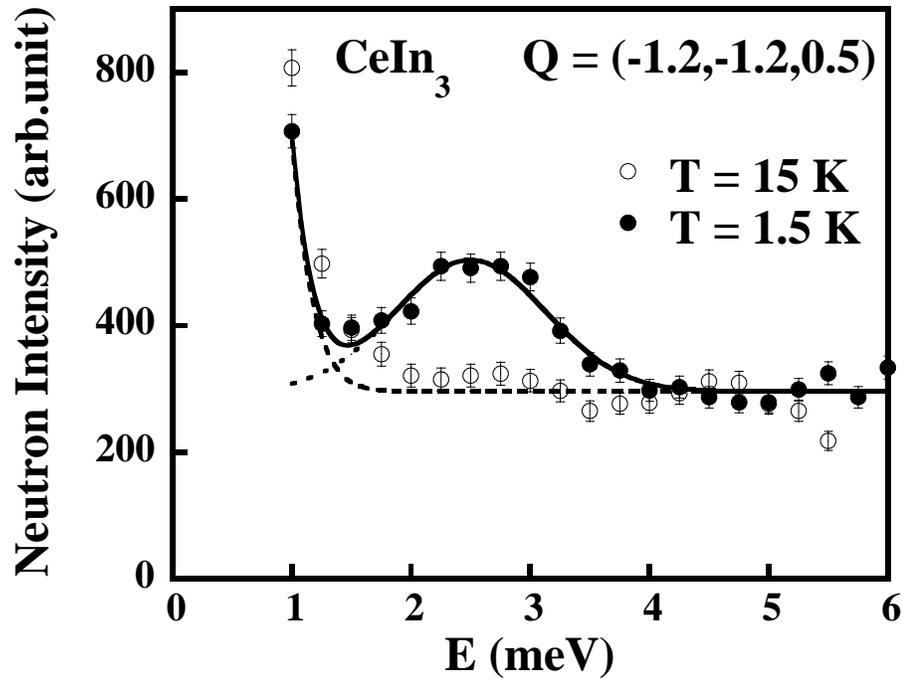,height=150mm,angle=0}
    \caption{Energy scans at $\textbf{Q}=(-1.2,-1.2,0.5)$ for
    temperatures of $T=1.5$ and 15 K obtained on the IN22 spectrometer
    using a final energy of 14.7 meV.}
    \label{fig3}
\end{figure}

\begin{figure}[h]
    \centering
    \epsfig{file=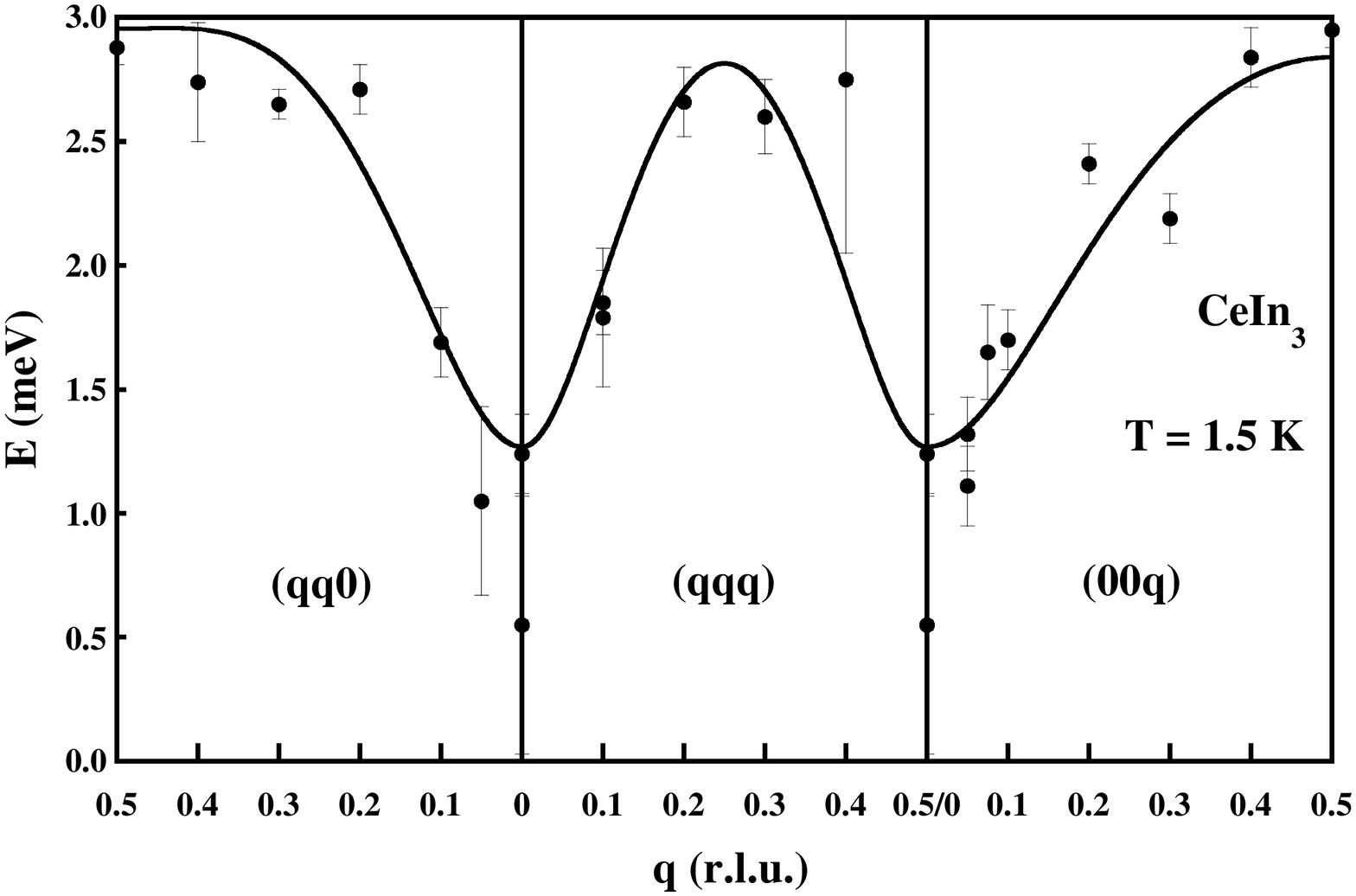,height=120mm,angle=0}
    \caption{Spin-wave dispersion of CeIn$_{3}$ along the three main
    directions $(0,0,q)$,
    $(q,q,0)$ and $(q,q,q)$ at $T=1.5$ K, obtained on the IN22
    spectrometer using a final energy of 14.7 meV.  The solid line is a fit obtained using the
    molecular-field random-phase approximation.}
    \label{fig4}
\end{figure}


\section*{References}
\begin{thebibliography}{<20>}

\bibitem{morin} Morin P., Vettier C., Flouquet J., Konczykowsky M., Lassailly Y., Mignot
 J.M. and Welp U. 1988 {\it J Low Temp Phys} {\bf 70} 377.
\bibitem{benoit} Benoit B., Boucherle J.X., Convert P., Flouquet J., Palleau J.
and Schweizer J. 1980 {\it Solid State Comm.} {\bf 34} 293.
\bibitem{mathur} Mathur N.D., Grosche F.M., Julian S.R., Walker I.R., Freye D.M.,
 Haselwimmer R.K.W. and Lonzarich G.G. 1998 {\it Nature} {\bf 394} 39.
\bibitem{knebel} Knebel G., Braithwaite D., Canfield P.C., Lapertot G., Flouquet J. 2001
 {\it Phys. Rev. B} {\bf 65} 24425.
\bibitem{millis} Millis A. 1993 {\it Phys. Rev. B} {\bf 48} 7183.
\bibitem{moriya} Moriya T. and Takimoto T. 1995 {\it J. Phys. Soc.
Japan} {\bf 64} 960.
\bibitem{thompson} Thompson J.D., Movshovich R., Fisk Z., Bouquet F.,
 Curro N.J., Fisher R.A., Hammel P.C., Hegger H., Hundley M.F., Jaime M.,
  Pagliuso P.G., Petrovic C., Phillips N.E. and Sarrao J.L. 2001 {\it
J. Magn. Magn. Mat.} {\bf 226-230} 5.
\bibitem{monthoux} Monthoux P. and Lonzarich G.G. 2001 {\it Phys. Rev. B} {\bf 63} 054529.
\bibitem{petrovic} Petrovic C., Pagliuso P.G., Hundley M.F., Movshovich R., Sarrao J.L.,
 Thompson J.D., Fisk Z., and Monthoux P. 2001 {\it J. Phys.: Condens. Matter} {\bf 13}
L337.
\bibitem{lawrence} Lawrence J.M. and Shapiro S.M. 1980 {\it Phys. Rev. B} {\bf
22} 4379.
\bibitem{murani} Murani A.P., Taylor A.D., Osborn R. and Bowden Z.A. 1993 {\it Phys. Rev. B}
{\bf 48} 10606.
\bibitem{canfield} Canfield P.C. and Fisk Z. 1992 {\it Philos. Mag. B}
{\bf 65} 1117.
\bibitem{lea} Lea K.R., Leask M.J.M., Wolf W.P. 1962 {\it J. Phys. Chem. Solids}
{\bf 23} 1381.
\bibitem{buschow} Buschow K.H.J., De Wijn H.W. and Van Diepen A.M. 1969 {\it J. Chem. Phys.}
{\bf 50} 137.
\bibitem{boucherle} Boucherle J.X., Flouquet J., Lassailly Y., Palleau J. and Schweizer J.
 1983 {\it J. Magn. Magn. Mat.} {\bf 31-34} 409.
\bibitem{buyers} Buyers W.J.L., Holden T.M. and Perreault A. 1975 {\it Phys. Rev. B}
{\bf 11} 266.
\bibitem{wang} Wang Y.L. and Cooper B.R. 1970 {\it Phys. Rev. B} {\bf
2} 2607.
\bibitem{gross} Gro\ss\,  W., Knorr K., Murani A.P. and Buschow K.H.J. 1980 {\it Z. Phys. B} {\bf 37} 123.
\bibitem{lassailly} Lassailly Y., Burke S.K. and Flouquet J. 1985
{\it J. Phys. C} {\bf 18} 5737.
\bibitem{kawasaki} Kawasaki S., Mito T., Zheng G.-q., Thessieu C., Kawasaki Y., Ishida K., Kitaoka Y.,
 Muramatsu T., Kobayashi T.C., Aoki D., Araki S., Haga Y., Settai R. and $\bar{O}$nuki Y. 2001 {\it Phys. Rev. B} {\bf
65} 020504.
\bibitem{hanzawa} Hanzawa K., Yamada K. and Yosida K. 1985 {\it J. Magn. Magn. Mat.} {\bf
47-48} 357.
\bibitem{us} Knafo W. et al., {\it in preparation}
\bibitem{kohori} Kohori Y., Kohara T., Yamato Y., Tomka G. and Riedi P.C. 2000
{\it Physica B} {\bf 281-282} 12.
\bibitem{vandijk} Van Dijk N.H., F{\aa}k B., Charvolin T., Lejay P. and Mignot J.M.
2000 {\it Phys. Rev. B} {\bf 61} 8922.
\bibitem{grier} Grier B.H., Lawrence J.M., Murgai V. and
Parks R.D. 1984 {\it Phys. Rev. B} {\bf 29} 2664.
\bibitem{demuer} Demuer A., Jaccard D., Sheikin I., Raymond S., Salce B., Thomasson J.,
 Braithwate D. and Flouquet J. 2001 {\it J. Phys. Condens. Matter} {\bf 13} 9335.
\bibitem{steeman} Steeman R.A., Mason T.E., Lin H., Buyers W.J.L.,
Menovsky A.A., Collins M.F., Frikkee E., Nieuwenhuys G.J. and
Mydosh J.A. 1990 {\it J. Appl. Phys.} {\bf 67} 5203.
\bibitem{grosche} Grosche F.M., Julian F.R., Mathur N.D. and
Lonzarich G.G. 1996 {\it Physica B} {\bf 223-224} 50.

\end{thebibliography}
\end{document}